\documentclass[journal,comsoc,twocolumn]{IEEEtran}
\usepackage{amsthm,amsmath,amssymb}
\usepackage{amsfonts,amssymb}
\usepackage[T1]{fontenc}
\usepackage{hyperref}
\usepackage{multirow}
\usepackage{diagbox}
\usepackage{cite}
\usepackage[pdftex]{graphicx}
\usepackage{graphicx, subfigure}
\usepackage{threeparttable}
\usepackage{color}
\usepackage{algorithm}
\usepackage{algorithmic}
\usepackage{tikz}
\usepackage{xcolor}

\hypersetup{
	colorlinks=true,
	linkcolor=black}

\begin{document}
\newcommand{\tabincell}[2]{\begin{tabular}{@{}#1@{}}#2\end{tabular}}
\renewcommand{\algorithmicrequire}{\textbf{Input:}}
\renewcommand{\algorithmicensure}{\textbf{Output:}}

\title{Deep Learning-based CSI Feedback for RIS-assisted Multi-user Systems}

\author{\normalsize {Jiajia~Guo, \IEEEmembership{\normalsize {Member,~IEEE}},
Xi~Yang,
Chao-Kai~Wen, \IEEEmembership{\normalsize {Fellow,~IEEE}},
Shi~Jin, \IEEEmembership{\normalsize {Fellow,~IEEE}},
and Geoffrey~Ye~Li, \IEEEmembership{\normalsize {Fellow,~IEEE}}
}

\thanks{J.~Guo, X~Yang, and S.~Jin are with the National Mobile Communications Research Laboratory, Southeast University, Nanjing, 210096, P. R. China (email: \{jiajiaguo, yangxi, jinshi\}@seu.edu.cn).}
\thanks{C.-K.~Wen is with the Institute of Communications Engineering, National Sun Yat-sen University, Kaohsiung 80424, Taiwan (e-mail: chaokai.wen@mail.nsysu.edu.tw).}
\thanks{G. Y. Li is with the Department of Electrical and Electronic Engineering,
Imperial College London, London, UK (e-mail: geoffrey.li@imperial.ac.uk).}
}

 \maketitle
\begin{abstract}
In the realm of reconfigurable intelligent surface (RIS)-assisted wireless communications, efficient channel state information (CSI) feedback is paramount. This paper introduces RIS-CoCsiNet, a novel deep learning-based framework designed to greatly enhance feedback efficiency. By leveraging the inherent correlation among proximate user equipments (UEs), our approach strategically categorizes RIS-UE CSI into shared and unique data sets. This nuanced understanding allows for significant reductions in feedback overhead, as the shared data is no longer redundantly relayed. Setting RIS-CoCsiNet apart from traditional autoencoder systems, we incorporate an additional decoder and a combination neural network at the base station. These enhancements are tasked with the precise retrieval and fusion of shared and individual data. And notably, all these innovations are achieved without modifying the UEs. For those UEs boasting multiple antennas, our design seamlessly integrates long short-term memory modules, capturing the intricate correlations between antennas. With a recognition of the non-sparse nature of the RIS-UE CSI phase, we pioneer two magnitude-dependent phase feedback strategies. These strategies adeptly weave in both statistical and real-time CSI magnitude data. The potency of RIS-CoCsiNet is further solidified through compelling simulation results drawn from two diverse channel datasets.
\end{abstract}


\section{Introduction}
\label{introduction}

\label{s1}

\IEEEPARstart{I}{n} June 2023, the International Telecommunication Union (ITU) approved the main framework and objectives for sixth-generation (6G) cellular networks \cite{ITU2030}. Six foundational usage scenarios for 6G have been identified: immersive communication, hyper-reliable low-latency communication, massive communication, ubiquitous connectivity, integrated AI and communication, and combined sensing and communication. Meeting these scenarios mandates advancements in 6G's peak data rate, user experience, coverage, sustainability, and more. In response these requirements, several new radio interface techniques, such as reconfigurable intelligent surfaces (RISs) \cite{9140329,9530717} and AI \cite{8663966,9040202}, are emerging. RISs, with their myriad controllable elements, reshape the propagation environment by tweaking each element's electromagnetic response (like phase, amplitude, and polarization \cite{9464298}) rather than merely adapting. This opens up new optimization paths for wireless systems, vital for achieving 6G's objectives.

\begin{figure}[t]
    \centering
    \subfigure [\label{systemModel1}RIS-assisted coverage enhancement.]{
     \includegraphics[width=0.3\textwidth]{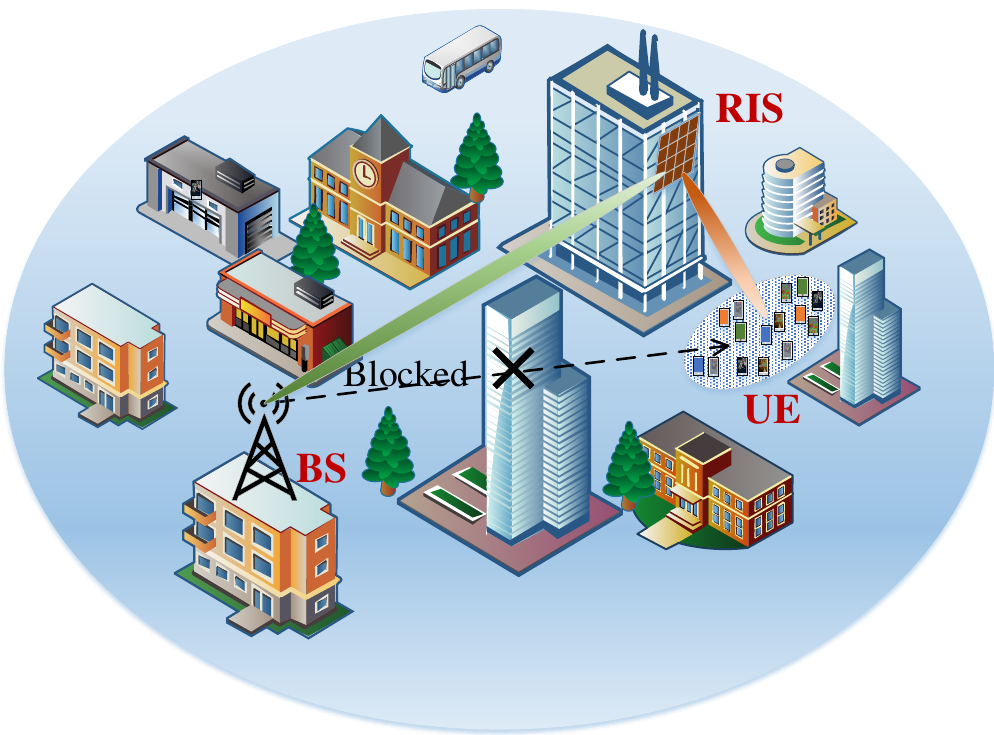}}
     \subfigure [\label{systemModel2}RIS-assisted capacity improvement.]{
     \includegraphics[width=0.3\textwidth]{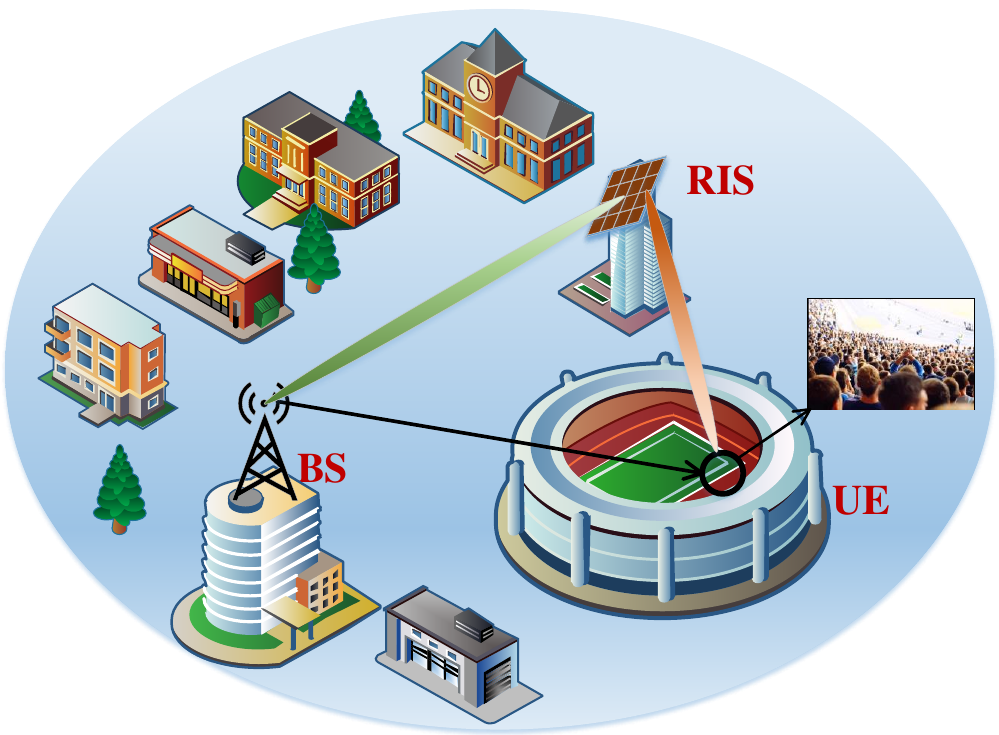}}
 \caption{\label{systemModel}Representation of two scenarios using RIS-enhanced communication.}
 \vspace{-0.8cm}
\end{figure}

However, these benefits manifest only with meticulous design of RIS element responses. Base station (BS) beamforming also demands careful design. Both passive and active beamforming designs \cite{8647620} depend on downlink channel state information (CSI). In time-division duplexing, channel reciprocity is used for CSI, but frequency-division duplexing (FDD) systems, more common in cellular networks, do not have this advantage. In FDD, the BS sends pilot symbols, allowing user equipment (UE) to estimate CSI, which is then communicated back to the BS. With the integration of RIS, 5G new radio (NR) systems using massive multiple-input multiple-output (MIMO) technology \cite{6798744} grapple with increased feedback overhead. For instance, in a scenario shown in Fig. \ref{systemModel1} where the direct BS-UE link is obstructed but an indirect BS-RIS-UE link exists, feedback overhead rises significantly, emphasizing the need for an efficient feedback system for RIS-supported communication.

Within the context of CSI feedback, one well-known promising technique is compressive sensing (CS) theory, which exploits the CSI sparsity in certain domains \cite{8284057}. In \cite{6214417}, the spatial-frequency domain's sparsity, resulting from the short distance between antennas, is used to compress CSI and reduce feedback overhead. A novel CS-based feedback algorithm, based on the sparsity of the cascaded BS-RIS-UE channel, is proposed in \cite{9497113}. While existing CS-based methods significantly reduce feedback overhead \cite{7174558}, the two main challenges persist. Firstly, the reconstruction from the compressed CSI at the BS is viewed as an optimization problem addressed by iterative algorithms, consuming considerable time and computational resources. Secondly, the existing methods exploits only sparsity but neglect the environment knowledge.

Deep learning (DL) has recently shown promise in addressing the CSI feedback challenges \cite{9931713}. The introduction of DL to CSI feedback was first seen in \cite{8322184} with the autoencoder model, where the encoder (UE) and decoder (BS) manage compression and reconstruction. This method focuses on capturing the predominant features of the input. By leveraging expert insights, like time correlation \cite{8482358}, and innovative neural network (NN) structures, such as the multi-resolution convolutional model \cite{lu2019multi}, CSI feedback accuracy improved. Recognizing its efficacy, the 3GPP spotlighted DL-based CSI feedback for 5G-Advanced studies \cite{techrep}. This technique has also been explored in RIS-assisted communications. In \cite{9592779}, the phase shift matrix from the UE to the RIS is conveyed via an autoencoder, augmented with a global attention mechanism \cite{yu2022pushing}. The Transformer model in \cite{9856664} compresses and extracts cascaded BS-RIS-UE CSI. Moreover, \cite{9969163} introduces a DL-based dual-timescale CSI feedback system, addressing both stable BS-RIS and dynamic RIS-UE channels \cite{9973349}. Yet, channel estimation in wholly passive RIS-assisted setups remains challenging. Some propose adding active elements to RIS, forging a hybrid system to simplify channel estimations \cite{9127834}. Given the unique feedback frequencies between BS-RIS and RIS-UE in RIS-backed communications \cite{9973349}, an optimized RIS-UE feedback algorithm is crucial.

It has been observed in several studies \cite{4699493,5285208,7905944,8552436,8471108,du2020shared} that proximate users' CSI often reveals shared multipath components and comparable sparsity patterns due to communal local scattering clusters. Such insights have bolstered CS-based channel estimation. However, in typical standard communication contexts, UEs are spread across vast areas, which naturally leads to reduced correlation between their CSIs. Fortunately, this scenario markedly changes in RIS-assisted communication use cases. Fig. \ref{systemModel} shows two RIS-assisted scenarios: coverage enhancement and capacity boost. In the former, direct transmission is hindered, necessitating RIS for extended coverage, concentrating UEs in a defined area. For example, \cite{9940989} employs RIS for UEs in an underground lot, implying significant channel correlation. Conversely, in the capacity-boost context depicted in Fig. \ref{systemModel2}, RIS boosts a specific area's capacity, like event-stadiums, emphasizing pronounced CSI correlation among UEs. Thus, harnessing nearby UEs' correlation is vital to reduce feedback overhead in RIS-supported channels.

In this paper, we introduce a DL-based CSI feedback framework named RIS-CoCsiNet for RIS-assisted multi-user systems. The primary objective of RIS-CoCsiNet is to harness the correlation among the RIS-UE channels of nearby UEs to decrease the feedback overhead inherent in RIS-assisted systems. Notably, within RIS-CoCsiNet, the RIS-UE CSI magnitude and phase are fed back separately. During the RIS-UE CSI magnitude feedback phase, the magnitude information is bifurcated into components shared among nearby UEs and components exclusive to individual UEs. The essence of exploiting this correlation is to minimize the overhead arising from redundantly feeding back shared information. An end-to-end approach at the BS seamlessly integrates these data types, all gleaned from the data. Conversely, the RIS-UE CSI phase, being non-sparse, poses challenges for compression, if not rendering it impossible. Notably, minuscule magnitude CSIs don't necessitate highly precise phase feedback. Drawing inspiration from \cite{8638509}, we propose two magnitude-dependent phase feedback (MDPF) NNs, wherein statistical and immediate magnitude information bolster the NN's phase feedback for the RIS-UE channel.
The key contributions of this paper include:
\begin{itemize}
\item {\bf Novel Cooperative Mechanism}: We introduce a cooperation mechanism into the autoencoder-based CSI feedback framework for RIS-assisted multi-user systems. This mechanism accounts for the correlation in the RIS-UE channels. Specifically, we split the RIS-UE CSI magnitude information into two parts: shared among nearby UEs and owned by individual UEs, as discussed in \cite{8552436}. An additional shared decoder at the BS is employed to recover shared information from nearby UE feedback bits.

\item {\bf LSTM-Enhanced Autoencoder Framework}: For UEs with multiple antennas, we enhance the BS decoder with Long Short-Term Memory (LSTM) architecture to better exploit the correlation among different antennas within the same UE.

\item {\bf Bitstream Formation Techniques}: We investigate and adopt two representative methods, quantization and binarization, for generating bitstreams in DL-based CSI feedback at the UE.

\item {\bf MDPF Approaches for RIS-UE CSI Phase}: To tackle the challenge of compressing the non-sparse RIS-UE CSI phase, we propose two MDPF frameworks. These frameworks incorporate statistical and instantaneous magnitude information into NNs.
\end{itemize}

While the primary focus of this paper is on leveraging the correlation among nearby UEs to optimize the CSI feedback overhead in RIS-assisted systems, our scope remains limited to the RIS-UE channel, employing only rudimentary NN architectures. Feedback for the BS-RIS channel over larger timescales and the BS-UE channel in RIS-assisted capacity enhancement scenarios can be achieved through a potent autoencoder, such as the Quan-Transformer described in \cite{9856664}. Implementing more sophisticated NN designs might offer enhanced performance, but this falls beyond our paper's purview.

The rest of this paper is organized as follows: Section \ref{s2} describes the RIS-assisted multi-user system model and the DL-based CSI feedback process. Section \ref{s4} details the cooperative RIS-UE CSI feedback approach, NN architecture, baseline framework, bitstream generation, and the CSI phase feedback method. Section \ref{s5} showcases numerical results, illustrating the RIS-CoCsiNet mechanism through parameter visualization. Section \ref{s6} concludes the paper.

\section{System Model} \label{s2}

%

\subsection{RIS-assisted Multi-user System Model}
\label{H-C}

We consider a RIS-assisted multi-user system where the BS and the UE are equipped with $N_{\rm t}$ (${\gg 1}$) antennas and $N_{\rm r}$ (${\ge 1}$) antennas, respectively. The RIS comprises $N_{\rm s}$ (${\gg 1}$) passive elements. This paper focuses on the RIS-assisted coverage enhancement scenario, in which there is no direct transmission link between the BS and the UE.
The downlink transmission signal model for the $k$th UE can be expressed as:
\begin{equation}
    {\bf y}_k = \widetilde{\bf H}_k {\bf \Phi}  {\bf G} {\bf w}_k { x}_k + {\bf z}_k,
\end{equation}
where ${{\bf y}_k\in \mathbb{C}^{N_{\rm r} \times 1}}$ is the received signal at the $k$th UE; ${\widetilde{\bf H}_k \in \mathbb{C}^{N_{\rm r} \times N_{\rm s}}}$ represents the RIS-UE channel between the BS and the $i$th UE; ${\bf G} \in \mathbb{C}^{N_{\rm s} \times N_{\rm t}}$ is the BS-RIS channel shared by all UEs; diagonal matrix ${\bf \Phi} = {\rm diag}(\phi_1,\ldots,\phi_{N_{\rm s}})\in \mathbb{C}^{N_{\rm s} \times N_{\rm s}}$ represents the RIS reflecting matrix; ${{\bf w}_k \in \mathbb{C}^{N_{\rm t} \times 1}}$ is the beamforming vector for the BS; $x_k$ is the downlink transmitted signal; and ${\bf z}_k \in \mathbb{C}^{N_{\rm r} \times 1}$ is the complex additive noise. Given the context, this work mainly considers feedback for the RIS-UE channel, $\widetilde{\bf H}_k$.

Based on the spatial multipath channel model \cite{doi:10.1002/ett.928}, the downlink RIS-UE channel matrix, $\widetilde{\bf H}_k$, is given by
\begin{equation}
\widetilde{\bf H}_k= \sum_{l=1}^{L} g_{k,l}\,{\bf a}_{\rm r}(\varphi_{{\rm r},k,l}) {\bf a}_{\rm s}^H(\varphi_{{\rm s},k,l}),
\end{equation}
where $L$ signifies the total path number; $g_{k,l}$ represents the complex gain of the $l$-th path; ${\bf a}_{\rm r}(\cdot)$ and ${\bf a}_{\rm s}(\cdot)$ denote the steering vectors for the receiver (UE) and transmitter (RIS), respectively; and $\varphi _{r,l}$ and $\varphi_{{\rm s},k,l}$ correspond to the angle-of-arrival (AoA) and the angle-of-departure (AoD).
Assuming both the RIS and the UE utilize uniform linear array (ULA) antennas (passive elements) for simplicity\footnote{The proposed can also be easily extended to other array topologies, such as uniform planar antenna array (UPA).}, the respective steering vectors are expressed as:
\begin{subequations}
\begin{align}
{\bf a}_{\rm r}(\varphi_{{\rm r},k,l}) &= {\left[1, e^{-j\kappa d\sin(\varphi_{{\rm r},k,l})}, \ldots, e^{-j \kappa d(N_{\rm r}-1)\sin(\varphi_{{\rm r},k,l})} \right]}^T, \\
{\bf a}_{\rm s}(\varphi_{{\rm s},k,l}) &= {\left[1, e^{-j\kappa d\sin(\varphi_{{\rm s},k,l})}, \ldots, e^{-j \kappa d(N_{\rm s}-1)\sin(\varphi_{{\rm s},k,l})} \right]}^T.
\end{align}
\end{subequations}
Here, ${\kappa = 2\pi/\lambda}$ denotes the wavenumber with $\lambda$ being the wavelength, and $d$ is the spacing between antenna elements.

Given the RIS's typical high placement and the limited RIS-UE channel path number $L$ due to minimal scatterers, the RIS-UE CSI matrix $\widetilde{\bf H}_k$ in the angle domain exhibits sparsity. This sparse matrix can be derived using the discrete Fourier transform (DFT):
\begin{equation}
\label{HDFT}
{\bf H}_k = {\bf F}_{\rm r}  \widetilde{\bf H}_k  {\bf F}_{\rm s} ,
\end{equation}
where ${\bf F}_{\rm r}$ and ${\bf F}_{\rm s}$ are ${N_{\rm r} \times N_{\rm r}}$ and ${N_{\rm s} \times N_{\rm s}}$ DFT matrices, respectively.

\begin{figure}[t]
    \centering
    \includegraphics[width=0.4\textwidth]{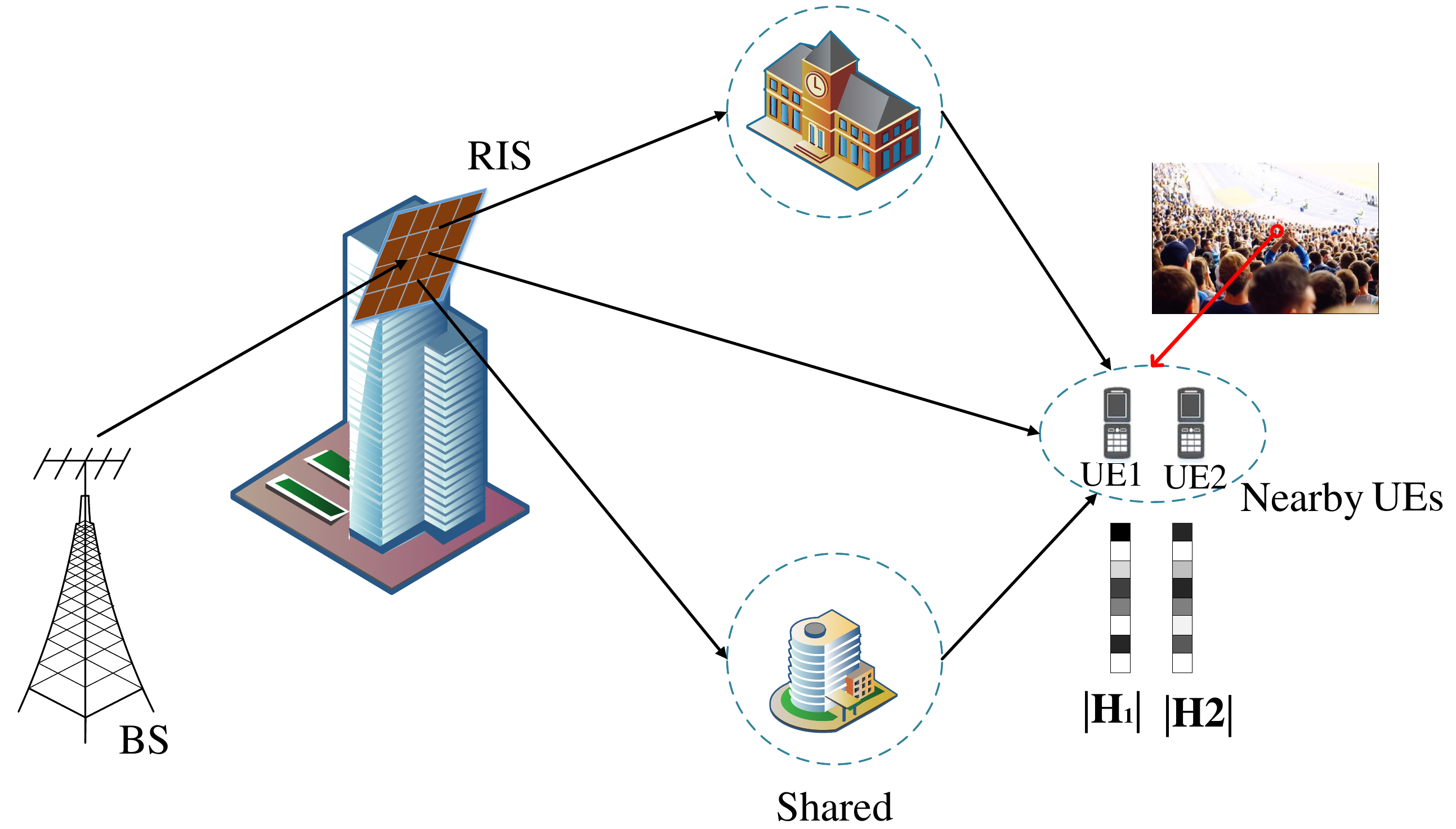}
    \caption{\label{ChannelModel}Depiction of an RIS-enhanced multi-user setup featuring two proximate UEs.}
    \vspace{-0.6cm}
\end{figure}

\subsection{Conventional DL-based Digital CSI Feedback Process}

We assume that the UE has estimated the CSI. After acquiring the channel matrix ${\bf H}_k$, UE employs an NN for CSI compression, producing measurement vectors in 32-bit floating point format. These vectors are then quantized, generating bitstreams that are promptly sent back to the BS. The BS then reconstructs the CSI, considering the quantization noise. This entire procedure is captured by:
\begin{equation}
\label{totalProcess}
\widehat{\bf H}_k =  f_{\rm de}({\sf D}({\sf Q}(f_{\rm en}({\bf H}_k;\,\Theta_{\rm en})));\,\Theta_{\rm de}) ,
\end{equation}
where $f_{\rm en}(\cdot)$ and $f_{\rm de}(\cdot)$ denote the DL-based compression (encoder) and decompression (decoder) functions of the UE and BS respectively. ${\sf Q}$ and ${\sf D}$ symbolize the quantization and dequantization operations\footnote{Dequantization involves losslessly reconstructing quantized measurement vectors using feedback bits.}. Parameters $\Theta_{\rm en}$ and $\Theta_{\rm de}$ correspond to the UE's NN-based encoder and the BS's decoder, respectively.

The total feedback overhead, specifically the bitstream length, is calculated as:
\begin{equation}
N_{\rm bits} = 2N_{\rm r} N_{\rm s} \times \gamma \times B,
\end{equation}
with $2N_{\rm r} N_{\rm s}$ being the RIS-UE CSI dimension, $\gamma$ the compression ratio, and $B$ the quantization bit count.

 \begin{figure*}[t]
    \centering
    \includegraphics[width=0.8\linewidth]{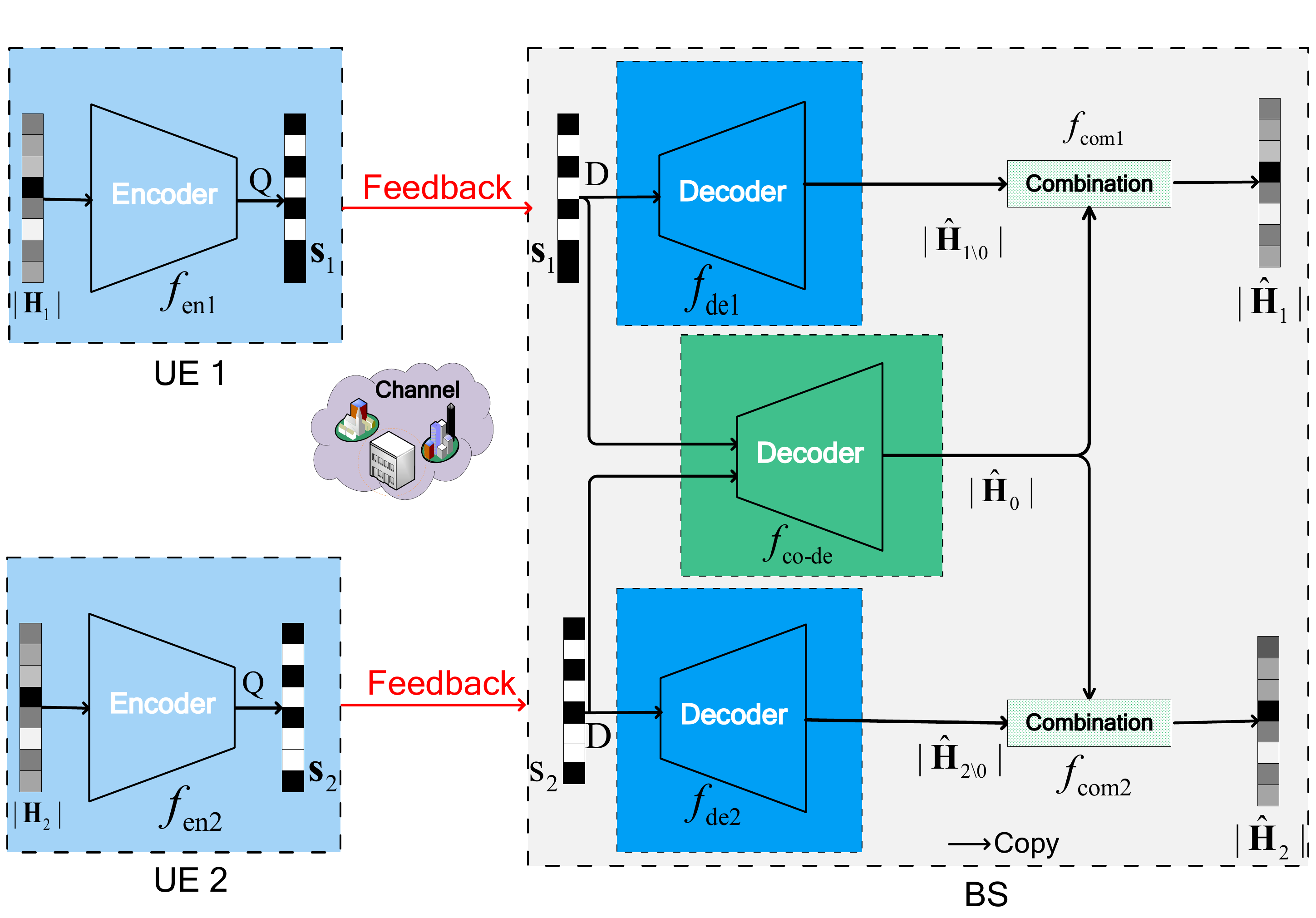}
    \caption{\label{cooperationFig}Schematic of the cooperative NN structure for two adjacent UEs, comprising two encoders and three decoders: two distinct and one shared decoder. The UEs' encoders produce the bitstream, which is then sent back through the uplink channel. At the BS, decoders utilize FC layers to extract shared RIS-UE CSI and specific information from the feedback bitstreams.}
    \vspace{-0.5cm}
\end{figure*}

Optimization of this procedure involves an end-to-end approach, refining the NN parameters $\{\Theta_{\rm en}, \Theta_{\rm de} \}$ to minimize the associated loss function. A prevalent loss function in CSI feedback is the mean-squared error (MSE):
\begin{equation}
\label{OptimizationGoal0}
 \min_{\Theta_{\rm en}, \, \Theta_{\rm de}}  {\left\|  {\bf H}_k-   f_{\rm de}({\sf D}({\sf Q}(f_{\rm en}({\bf H}_k;\,\Theta_{\rm en})));\,\Theta_{\rm de})     \right\|}_2^2,
\end{equation}
where ${\Vert \cdot \Vert}_2$ symbolizes the Euclidean norm.
To compare the original and reconstructed RIS-UE CSI at the BS, the normalized MSE (NMSE) metric is applied:
\begin{equation}
{\rm NMSE} = \mathbb{E} {\left \{  \frac{  { \Vert  \widehat{\bf H}_k - {\bf H}_k \Vert }_2^2  }{ { \Vert   {\bf H}_k \Vert }_2^2 }   \right \}},
\end{equation}
where $\mathbb{E}\{\cdot\}$ denotes the average taken over all testing samples.

\section{DL-based CSI Feedback for RIS-assisted Multi-user Systems}
\label{s4}
In this section, we first discuss the RIS-assisted multi-user scenario where two adjacent UEs share specific reflection or diffraction mediums and have correlated CSI magnitudes. Unlike the conventional cooperation feedback strategy \cite{7933248} enabled by device-to-device communications, our DL-based cooperation strategy for RIS-UE channel feedback eliminates the need for additional operations at the UE. The main goal is to reduce the overhead stemming from repetitive feedback of RIS-UE CSI information shared by these adjacent UEs. We then detail the NN modules in RIS-CoCsiNet.

\subsection{Cooperation Strategy}

\subsubsection{Framework}

Fig. \ref{ChannelModel} illustrates that UEs serviced by the RIS may share identical scatterers and similar deterministic multipath components when positioned in close proximity. As supported by \cite{4699493,8552436,995511,104090,du2020shared}, the channel parameters of these adjacent UEs are akin, with the magnitude of RIS-UE CSI in the angular domain showing strong correlation. No correlation is observed in the RIS-UE CSI phase. Additionally, the study in \cite{8552436} leveraged the shared common sparsity structures within the RIS-UE CSI, dividing the sparse vector of CSI into commonly shared and individual sparse representation vectors. Drawing inspiration from these observations, the RIS-UE CSI magnitude, denoted as $|{\bf H}_k|$, for the $k$-th UE is split into two sections: one shared by adjacent UEs and one uniquely owned by the individual UE.

To leverage this characteristic, the prevalent autoencoder-based NN framework requires modifications. However, in view of the UE's constrained computational capabilities and storage, the encoder framework of the UE remains as is---maintaining its NN architecture and parameter count. Therefore, the feedback bitstreams for the two UEs can be expressed as:
\begin{subequations}
\begin{align}
    {\bf s}_1 &= {\sf Q}\big(f_{\rm en1}(|{\bf H}_1|;\,\Theta_{\rm en1})\big), \\
    {\bf s}_2 &= {\sf Q}\big(f_{\rm en2}(|{\bf H}_2|;\,\Theta_{\rm en2})\big),
\end{align}
\end{subequations}
where $f_{\rm en1} (\cdot)$ and $f_{\rm en2} (\cdot)$ are the encoder modules at the two UEs, and $\Theta_{\rm en1}$ and $\Theta_{\rm en2}$ are the NN parameters of the encoder module.
During training, two encoders are trained together to cooperatively compress the RIS-UE CSI. The mechanism will be given in Fig. \ref{FCFigD2D}.

Alterations are confined to the NN framework at the BS. There, the decoder encompasses modules dedicated to retrieving RIS-UE CSI vectors with both shared and individual components. For two adjacent UEs, the BS features a trio of decoder modules: a shared decoder and two distinct decoders, depicted in Fig. \ref{cooperationFig}. Since the shared information amongst nearby UEs remains consistent, the shared decoder module can serve both UEs to reconstitute the shared information $|\widehat{\bf H}_{0}|$ as:
\begin{equation}
|\widehat{\bf H}_{0}| = f_{\textrm{co-de}}\big({\sf D}({\bf s}_1),{\sf D}({\bf s}_2);\,\Theta_{\rm co-de}\big) ,
\end{equation}
where $\Theta_{\rm co-de}$ is the parameter of shared decoder $f_{\textrm{co-de}}(\cdot)$. Meanwhile, individual decoders work to reconstruct respective CSI information, $|\widehat{\bf H}_{1\setminus 0}|$ and $|\widehat{\bf H}_{2\setminus 0}|$. Two subsequent modules then amalgamate the shared and individual UE information utilizing fully connected (FC) layers as depicted in:
\begin{subequations}
\begin{align}
|\widehat{\bf H}_1| &=f_{\rm com1}(|\widehat{\bf H}_{1\setminus 0}|,|\widehat{\bf H}_{0}| ;\,\Theta_{\rm com1}), \label{h1} \\
|\widehat{\bf H}_2| &=f_{\rm com2}(|\widehat{\bf H}_{2\setminus 0}|,|\widehat{\bf H}_{0}| ;\,\Theta_{\rm com2}). \label{h2}
\end{align}
\end{subequations}
Here, $f_{\rm com1} (\cdot)$ and $f_{\rm com2}(\cdot)$ are the combination modules at the two nearby UEs, and those $\Theta_{\rm com1}$ and $\Theta_{\rm com2}$ are the NN parameters of these two modules.
Finally, $|\widehat{\bf H}_1|$ and $|\widehat{\bf H}_2|$ represent the comprehensively reconstructed magnitude matrices of the RIS-UE CSI.

\subsubsection{Architecture}
Fig. \ref{cooperationFig} depicts the whole cooperative feedback framework, termed RIS-CoCsiNet, designed for the RIS-UE channels in RIS-assisted multi-user scenarios. This structure encompasses encoders situated at the UE and decoders positioned at the BS.

Within the UE domain, the dual encoders produce bits through the FC layers and a binary (or quantization) layer. Once bitstream creation, these bits are fed back via uplink transmission. Concurrently, the shared decoder located at the BS extracts the communal CSI data $|\widehat{\bf H}_{0}|$ from the feedback received from the two proximate UEs. Simultaneously, dedicated decoders retrieve individual CSI data $|\widehat{\bf H}_{1\setminus 0}|$ and $|\widehat{\bf H}_{2\setminus 0}|$ pertinent to each of the two UEs. The conclusive CSI magnitude matrix emerges as an amalgamation of both the retrieved shared and unique CSI magnitude details. Notably, the co-decoder's network architecture mirrors that of its individual counterpart, with the distinction that the co-decoder's input ingests feedback from all collaborating UEs. The combination module is constituted by a simple NN.

The proposed RIS-CoCsiNet framework integrates both explicit and implicit cooperative mechanisms. Explicit cooperation manifests when the BS's decoder collectively extracts the RIS-UE CSI based on feedback from nearby UEs. Implicit cooperation, on the other hand, takes place during compression at the encoders. Contrasting with cooperation-absent NNs, which necessitate repetitive feedback of shared RIS-UE CSI information by the UE, the encoders within RIS-CoCsiNet collaboratively relay this shared RIS-UE CSI data, optimizing the process through end-to-end learning.

\subsubsection{Loss Function}

For effective training, both proximate UEs should undergo simultaneous training. This process ensures that RIS-CoCsiNet is adept at extracting and restoring the shared RIS-UE CSI data. As such, the proposed RIS-CoCsiNet is trained directly through a one-step end-to-end learning strategy. The training loss function represents the combined MSE of both UEs, given by:
\begin{equation}
\label{OptimizationGoalCoMa}
 {\cal L}_{\textrm{RIS-CoCsiNet}} = \left\| |{\bf H}_1|-|\widehat{\bf H}_1| \right \|_2^2  + \left \|  |{\bf H}_2|-|\widehat{\bf H}_2|  \right \|_2^2.
\end{equation}

\subsubsection{Complexity and Extension to Multiple UEs}
The encoder in RIS-CoCsiNet at the UE mirrors many of the prevailing feedback frameworks that do not incorporate UE cooperation. As such, the number of NN parameters and the floating-point operations remain unchanged. Although an additional shared decoder is present at the BS, heightening the NN's complexity, it marginally impacts the speed of CSI reconstruction. This is because the BS is not constrained by computational power, and the three decoders can be performed parallelly.

Furthermore, RIS-CoCsiNet is adaptable and can seamlessly be expanded to accommodate scenarios with multiple UEs. Every UE retains its encoder. If we assume the number of adjacent UEs to be $K$, then, the shared decoder at the BS will decipher shared data derived from feedback from all $K$ UEs. Correspondingly, $K$ distinct decoders will discern their individual data. Ultimately, combination modules amalgamate the shared and individual data to render the final CSI for all $K$ UEs.

\begin{figure*}[t]
    \centering
    \includegraphics[width=0.7\linewidth]{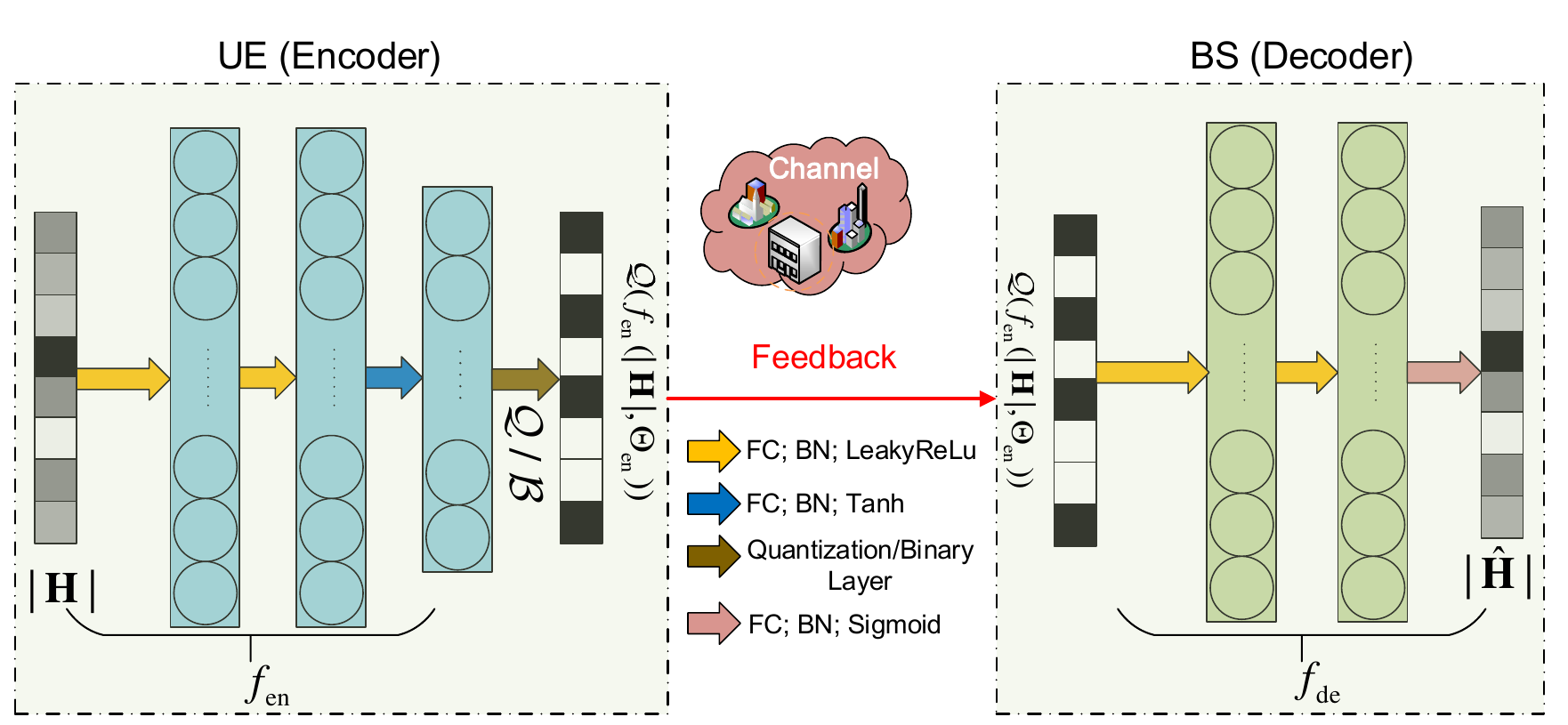}
    \caption{\label{NNarchitecture1}Diagram of the NN architecture comprising the encoder and decoder. The encoder at the UE forms the bitstream using the FC layers and a quantization/binary layer, then relays this bitstream through the uplink channel. The BS decoder retrieves the RIS-UE CSI from the feedback bitstreams utilizing the FC layers.}
    \vspace{-0.5cm}
\end{figure*}

\subsection{Modules within RIS-CoCsiNet}
In the following subsection, we begin by outlining the foundational NN architecture employed in this study. Subsequently, we delve into the binarization process, which is responsible for producing the bitstreams. This process can be viewed as an evolved form of one-bit quantization. To conclude, we introduce a phase strategy known as MDPF.

\subsubsection{Baseline NN Architecture}
\label{NNArch}

The foundational architecture of RIS-CoCsiNet is displayed in Fig. \ref{NNarchitecture1}, which is comprised of both an encoder and a decoder component.
The encoder situated at the UE consists of a series of three FC layers, which are subsequently followed by three batch normalization (BN) layers. The activation function for the first pair of BN layers is designated as the LeakyReLU function. Distinctively, this function allows for a non-zero gradient when the unit is not active, distinguishing it from the standard ReLU activation function. It can be mathematically expressed as:
\begin{equation}
{\rm LeakyReLu}(x)=
\begin{cases}
x, &{ \quad  x \ge 0},\\
a\cdot x, &{ \quad  x <0},
\end{cases}
\end{equation}
where $a$ is a negligible coefficient. The activation function allocated to the final BN layer is the hyperbolic tangent function $\tanh(\cdot)$. This function's objective is to yield values that lie within the continuous interval from $-1$ to $1$. As a result, the encoder's terminal layer can be identified as the quantization or binarization layer that has been discussed earlier.


In contrast, the decoder at the BS incorporates three FC layers\footnote{Should we opt for the quantization layer, a dequantizer becomes imperative; however, it is omitted from the current depiction.}. A distinction can be drawn between the NN layers of the encoder and the decoder. Specifically, the final FC layer of the decoder comprises $N_{\rm r} \times N_{\rm s}$ neurons, and the last BN layer utilizes the Sigmoid function to constrain outputs within the range of $[0,1]$.

For scenarios where the UE is outfitted with multiple antennas, the decoder integrates an additional LSTM architecture at its tail-end. This augmentation aids in harnessing the inherent correlation among the diverse antennas present within the same UE, akin to the design found in \cite{8482358}. The LSTM architecture has gained popularity for applications in time-fluctuating contexts due to its aptitude in extracting and capitalizing on time-based correlations, evident from instances like CsiNet-LSTM \cite{8482358}. Given its design for sequential data, LSTMs are adept at recognizing and leveraging correlations found in sequential ``frames,'' encompassing time, frequency, and spatial dimensions. Consequently, the LSTM's incorporation in this research is to discern and make use of the correlations exhibited by proximate antennas. The supplementary LSTM within the decoder exhibits exemplary CSI reconstruction capabilities, courtesy of its intrinsic memory cells that retain previously extracted information over extended durations for future use. An in-depth exploration of the LSTM module is available in \cite{8482358}.

\subsubsection{Binarization Operation}
Traditional DL-based methods, which can be interpreted as DL-based CS algorithms, encompass two primary operations: compression and quantization. From a DL standpoint, the first operation endeavors to produce latent vectors representing the original CSI or to extract the original CSI's features. The purpose of the quantization operation is to reduce feedback overhead by quantizing these features.

Moreover, binary representation facilitates the generation of bitstreams \cite{8941111}. To some degree, this binary representation-based CSI feedback can be viewed as a codebook-driven, data-dependent feedback strategy. Diverging from earlier approaches to codebook design, the NN-based encoder, inclusive of the binary layer, can be perceived as the module producing the codebook indices. These indices consist of a set of $\{-1,1\}$. In tandem, the NN-based decoder restores the CSI utilizing these codebook indices. According to \cite{raiko2014techniques}, the binarization operation can be divided into two key phases:
\begin{enumerate}
\item The neuron count in the encoder's last FC layer is adjusted to match the desired length of the feedback bits. To produce numbers within the continuous range $[-1,1]$, the activation function for the BN succeeding this FC layer is set as $\tanh(\cdot)$.

\item This real-valued representation then serves as input, and is transformed to produce a discrete output from the set $\{-1,1\}$. This transformation is achieved using the binarization function defined as:
\begin{equation}
b(x) = x + \epsilon,
\end{equation}
for $x \in [-1,1]$, where
\begin{equation}
\label{quantizationNoise}
\epsilon =
\begin{cases}
1-x,&{\rm{with \quad probability} \quad  \frac{1+x}{2}},\\
-1-x,&{\rm{with \quad probability} \quad  \frac{1-x}{2}},
\end{cases}
\end{equation}
and $\epsilon$ stands for the quantization noise with zero mean.
\end{enumerate}
Considering the nondifferentiable nature of the second step, its gradient is designated as one. This facilitates backpropagation through the operation, specifically through the binary layer, denoted as $\mathcal{B}$.

\begin{figure*}[t]
    \centering
    \subfigure [\label{NNarchitecture4Phase}Schematic of MDPF-1]{
     \includegraphics[scale=0.75]{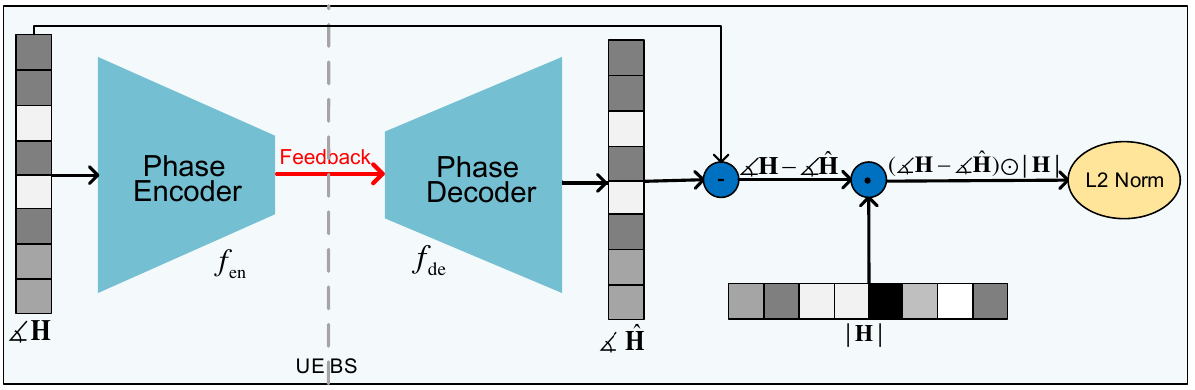}}
	\subfigure[\label{NNarchitecture4Phase2}Schematic of MDPF-2]{
	\includegraphics[scale=0.76]{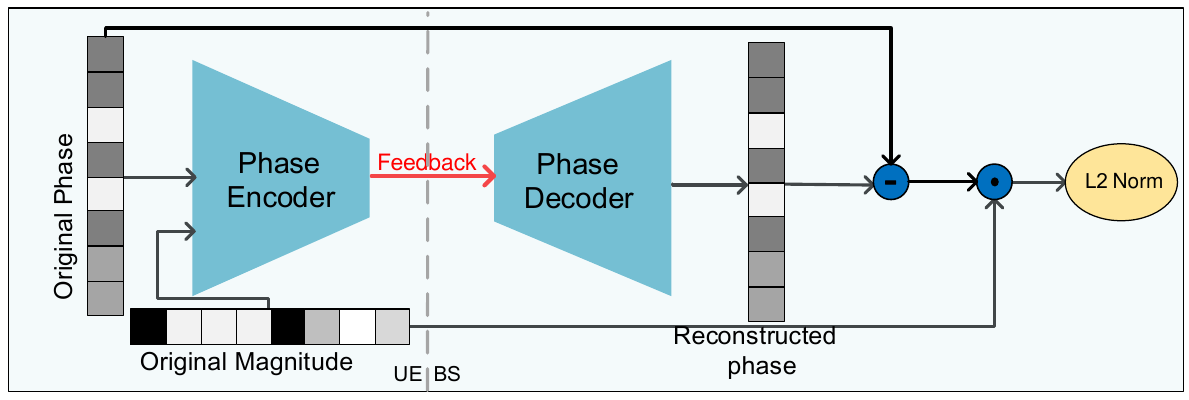}	}
	\caption{Schematics of MDPF-1 and MDPF-2. In MDPF-1, during training, the loss function is a weighted MSE, influenced by the RIS-UE CSI magnitude. This introduces statistical RIS-UE CSI magnitude data to the NNs. The quantization module post-encoder is omitted in this depiction. In MDPF-2, the RIS-UE CSI magnitude is fed into the encoder, and the final loss function for the RIS-UE CSI phase feedback NNs is the weighted MSE, contingent on the immediate CSI magnitude.}
\end{figure*}

\subsubsection{NN for Phase Feedback}
\label{loss}
In contrast to the RIS-UE CSI magnitude, which exhibits sparsity due to the presence of limited paths, the phase vector or matrix of the RIS-UE CSI lacks such sparsity. Attempting to relay phase information in the same manner as the magnitude feedback would result in the transmission of a plethora of extraneous data. Feedback bits associated with the RIS-UE CSI that have insignificant magnitude convey minimal valuable information, leading to resource wastage. Eliminating this superfluous information during the feedback process ensures optimized utilization of feedback overhead, significantly enhancing feedback performance.

The study in \cite{8638509} introduces a magnitude-dependent phase quantization strategy. In this approach, CSI coefficients with higher magnitudes undergo finer phase quantization and vice versa. This means the more crucial the CSI phase is, the more bits are allocated for it. Similarly, \cite{liu2019efficient} presents the use of an additional NN to determine the allocation of quantization bits.
Drawing inspiration from these studies, we adopt two magnitude-dependent strategies for CSI phase feedback. The differentiation between the two lies in their dependence: one strategy relies on the statistical RIS-UE CSI magnitudes, while the other is based on the instantaneous RIS-UE CSI magnitudes.

\paragraph{MDPF-1}
Rather than merely focusing on the allocation of quantization bits, as seen in prior research, the NN in the MDPF-1 strategy emphasizes coefficients with higher magnitudes by refining the loss function. In the magnitude feedback discussed earlier, the employed loss function is the straightforward MSE. This standard MSE loss function treats each coefficient in the matrix equally, striving to minimize the discrepancy between the actual values and their corresponding predictions. However, this simplistic approach fails to account for the relative importance of each phase coefficient.

To address this, as demonstrated in Fig. \ref{NNarchitecture4Phase}, the loss function is enhanced by incorporating critical data---the statistical magnitude information---as follows:
\begin{equation}
\label{OptimizationGoalPhase}
{\cal L}_{\rm phase} =\left\|  ( \measuredangle{\bf H} -   \measuredangle \widehat{\bf H}  ) \odot|{\bf H}|  \right\|_2^2.
\end{equation}
Here, $\measuredangle{\bf H}$ signifies the original phase of the RIS-UE CSI in radians, $\measuredangle\widehat{\bf H}$ denotes the predicted phase, $|{\bf H}|$ represents the associated RIS-UE CSI magnitude, and $\odot$ symbolizes the Hadamard product.

The NNs employed within the MDPF-1 approach echo the foundational NN architecture seen in magnitude feedback. During the parameter updating phase, larger discrepancies carry a significant weight on the MSE loss since the MSE amplifies these discrepancies by squaring them prior to averaging. Recognizing that the significance of a phase is determined by its corresponding CSI magnitude, a higher magnitude leads to an augmented MSE. Incorporating magnitude information into the phase feedback's loss function ensures that updates prioritize phase coefficients with substantial magnitudes. As the instantaneous CSI magnitude is not a direct input for this network during testing, our proposed methodology compresses and relays the CSI phase leveraging the statistical RIS-UE CSI magnitude data. This knowledge is acquired during training by integrating the instantaneous CSI magnitude into the loss function.

\paragraph{MDPF-2}

In contrast to MDPF-1, which does not utilize the instantaneous CSI magnitude as input for the phase feedback NN and solely relies on the statistical CSI magnitude information, MDPF-2 incorporates the instantaneous CSI magnitude directly into the encoder, as illustrated in Fig. \ref{NNarchitecture4Phase2}. All other components, including the loss function, NN architecture, and training strategy, remain consistent with those used in MDPF-1.

\section{Simulation Results and Discussions}
\label{s5}

This section provides a comprehensive numerical simulation to assess the feedback performance of the proposed feedback techniques, especially in RIS-assisted multi-user settings where UEs are in close proximity. We begin by detailing the simulation environment, including channel parameters, the nuances of the NN architecture, and specifics regarding NN training. Subsequently, we compare the outcomes of quantization and binarization. Through these simulations, we aim to highlight the effectiveness of the methodologies proposed in Section \ref{s4}, particularly the additional LSTM and the MDPF strategy. Finally, we compare the proposed cooperative NNs (termed as RIS-CoCsiNet) against those that overlook the RIS-UE CSI correlation among nearby UEs. Drawing from \cite{guo2019convolutional}, we further elucidate cooperative RIS-UE CSI feedback through visualizing NN parameters.

\subsection{Parameter Setting}
\subsubsection{RIS-UE Channel Generation}
In order to critically assess the efficacy of the proposed RIS-CoCsiNet, we leverage two distinct channel datasets. The first set is self-created, while the second is sourced from publicly accessible software. Taking heed from Section \ref{s1}, our focus remains strictly on the RIS-UE channel. Consequently, our simulations employ solely the RIS-UE channels to both train and test the proposed RIS-CoCsiNet.

The first method for generating RIS-UE channels is based on the 3GPP spatial channel model (SCM) \cite{doi:10.1002/ett.928}. Keeping with the parameters from \cite{8552436}, the downlink frequency is fixed at 2.17 GHz. The gap between individual antennas (or RIS elements) is half the wavelength, i.e., $d = c/(2f_0)$, where $c$ signifies the speed of light and $f_0$ stands at 2 GHz, representing the carrier frequency. There exist $L = 3,4,5,6$ random scattering clusters (paths) spanning from $- \pi/2$ to $\pi /2$. Configurations of RIS elements are either 64 or 256, while the UE antennas vary between 1 and 4. Taking into account the proximity of the UEs, such as in a stadium scenario, nearby UE's CSI often share a majority of the scattering clusters and exhibit similar path losses. The CSI parameters of the first UE, like AoA, AoD, and gain, are initialized with random values. For the second UE, these parameters are derived by appending small random variations to those of the first UE. This methodology yielded a substantial 100,000 pairs of CSI samples using MATLAB.

In contrast to the initial database creation strategy, the second dataset is produced through QuaDRiGa software \cite{QuaDRiGa}, ensuring compliance with 3GPP TR 38.901 v15.0.0 \cite{3gpp2018study}. The RIS, equipped with 256 elements, stands at a height of 25 m. A 20-story building is situated 100 m away from the RIS. Within this context, the RIS's primary function is to bolster communication capabilities for the UEs housed inside the building. These UEs are dispersed across a $\rm 20 \, m \times 20 \, m$ region, randomly assigned to any of the 20 floors. Each floor spans 3 m in height, with the UEs positioned at a height of 1.5 m. A combined 10,000 sets of CSI are randomly produced. Citing \cite{6gwhite}, we anticipate the device density to surge to hundreds per cubic meter in the foreseeable future, resulting in users being mere tens of centimeters apart. Consequently, every set comprises either two or four UEs, ensuring the distance between any two remains under 0.3 m.

Datasets are systematically partitioned into training (70\%), validation (10\%), and test (20\%) subsets. The initial dataset plays a pivotal role in updating the NN parameters, while the latter aids in circumventing over-fitting. As a precaution against overfitting, we have incorporated early stopping; it becomes operational as soon as performance metrics on the validation dataset begin to wane. Additionally, we have employed the bit per dimension (BPD) metric to gauge feedback overhead, a measure that is widely recognized in the realm of image compression. It is mathematically represented as:
\begin{equation}
{\rm BPD} = \frac{N_{\rm bits}}{N_{\rm r} N_{\rm s}}.
\end{equation}

\subsubsection{NN Architecture}
As briefly touched upon in Section \ref{NNArch}, the architecture of the NN is presented. We have confined our endeavors to the primary objective of this work, as opposed to the comprehensive design of NN architectures specifically for CSI feedback. Hence, only commonly used layers, namely FC, BN, and LSTM are employed in our setup. The architecture comprises both UE and BS components, each consisting of three FC layers. Given the robust computational prowess at the disposal of the BS, its FC layers are designed to be twice as wide as those of the UE. A comprehensive breakdown of the architectural specifics of the proposed NNs, including neuron count within FC layers and the associated activation functions, can be found in Table \ref{NNArchitecture}.

\begin{table}[t]
\caption{Detailed NN architecture.}
\label{NNArchitecture}
\centering
\begin{threeparttable}
\begin{tabular}{c|ccc}
\hline \hline
&Layer name& Output size& Activation function \\
\hline
\multirow{6}{*}{\rotatebox{90}{Encoder (UE)}} &  Input Layer& $N_{\rm r} \times  N_{\rm s} $ &  None\\
&Reshape1& $N_{\rm r}N_{\rm s} \times 1$ &  None\\
&FC1+BN1& $2N_{\rm r} N_{\rm s} \times 1$ &LeakyReLu\\
&FC2+BN2& $2N_{\rm r} N_{\rm s} \times 1$ &  LeakyReLu\\
&FC3+BN3& $N_{\rm {bits}} /B$  \tnote{*} or $N_{\rm {bits}} \times 1 $ &  tanh\\
&\tabincell{c}{Quantization\\or Binary layer}  &$ N_{\rm {bits}} \times 1 $&  None\\
\hline \hline
\multirow{7}{*}{\rotatebox{90}{Decoder (BS)}} & FC4+BN4& $4N_{\rm r} N_{\rm s} \times 1$ &LeakyReLu\\
&FC5+BN5& $4N_{\rm r} N_{\rm s} \times 1$ &  LeakyReLu\\
&FC6+BN6& $N_{\rm r} N_{\rm s} \times 1$ &  Sigmoid / tanh \tnote{**} \\
&Reshape2& $N_{\rm r}\times N_{\rm s} $ &  None\\
& LSTM1 &  $N_{\rm r}\times N_{\rm s} $ & -\\
& LSTM2 &  $N_{\rm r}\times N_{\rm s} $ & -\\
& LSTM3 &  $N_{\rm r}\times N_{\rm s} $ & -\\
\hline \hline
\end{tabular}
 \begin{tablenotes}
        \footnotesize
        \item[*] $B$ represents the number of quantization bits.
        \item[**] When the decoder is tasked with recovering the CSI magnitude, the Sigmoid function is selected. Conversely, for RIS-UE CSI phase recovery, the Tanh function is employed.
 \end{tablenotes}
\end{threeparttable}
\vspace{-0.6cm}
\end{table}

For the execution of simulations, we utilized TensorLayer version 1.11.0 and ran the operations on a NVIDIA DGX-1 workstation. Standard hyperparameters were set with a batch size of 200, a learning rate of 1e-3, and a maximum epoch count of 1,000. The optimization of the loss function during training leverages the Adam algorithm, which has become a staple in deep learning due to its efficiency and robustness.

\subsection{Performance of Baseline RIS-UE CSI Magnitude Feedback}
Let us delve into a comprehensive evaluation of the baseline NN-based feedback mechanism for RIS-UE CSI magnitude. For this assessment, the SCM dataset is utilized.

\begin{figure}[t]
    \centering
	\includegraphics[width=0.9\linewidth]{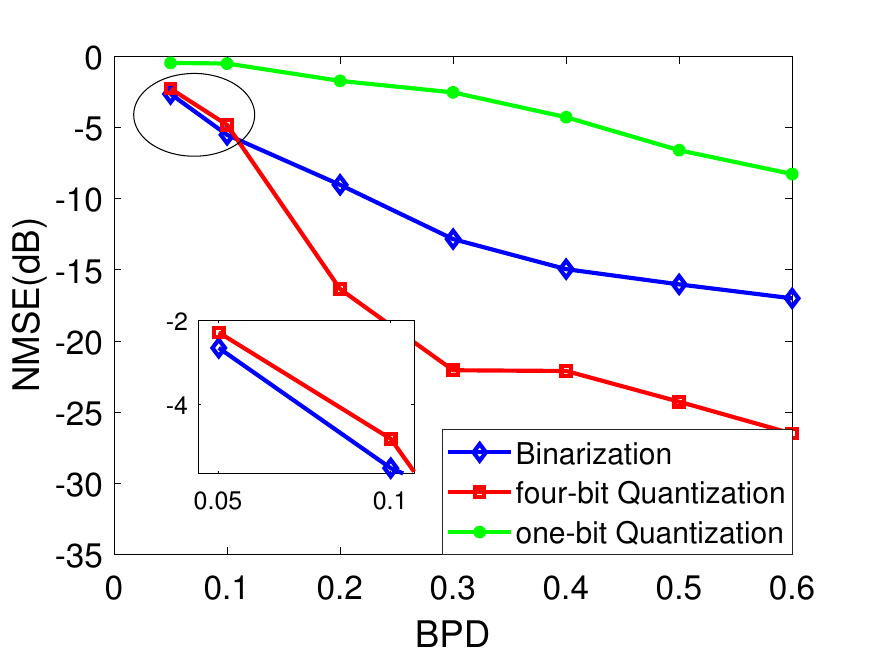}	
	\caption{\label{quantizationFig}
	NMSE (dB) performance comparison of RIS-UE CSI magnitude feedback between four and one-bit quantization, and binarization operation under different BPD. $N$ and $N_{\rm r}$ are set as 256 and 1, respectively. The first SCM dataset is adopted, and the channel path number is $L=5$.}
\end{figure}

To kick off, we distinguish between two primary methods used to generate bitstreams: quantization and binary representation. We set the stage with a single-user scenario, wherein the RIS is outfitted with 256 elements, and the UE features a singular antenna. The quantization bit number, denoted by $B$, is fixed at 4, a value justified as optimal from simulations showcased in \cite{8845636}. Fig. \ref{quantizationFig} portrays the BPD-NMSE trade-offs as accomplished by both quantization and binarization processes.

From one perspective, binarization can be construed as a subset of quantization when $B=1$. However, there is a stark difference in their operational dynamics. Unlike generic quantization, as exemplified in (\ref{quantizationNoise}), the binarization procedure incorporates a quantization noise, $\epsilon$, that exhibits a zero-mean trait. This ensures that the binary function $b(x)$ remains an unbiased estimator for the soft value $x$ --- mathematically, this translates to $\mathbb{E}_{\epsilon}\{ b(x)|x \} = x$ \cite{8941111}. A comparative analysis between binarization and one-bit quantization is showcased in Fig. \ref{quantizationFig}. The former evidently supersedes the latter by a considerable margin, underlining the pivotal role of a zero-mean $\epsilon$. In most scenarios, the performance via quantization remains superior to binarization, except in cases where the BPD is extraordinarily low, implying a strict constraint on feedback overhead.

\begin{table}[b]
\caption{Comparison of NMSE (in $\rm dB$) between FCNN and LSTM.}
\label{MAFig}
\centering
\begin{threeparttable}
\begin{tabular}{c|cccc}
\hline \hline
$L$ & 3      & 4      & 5      & 6      \\ \hline
LSTM  & -24.84 & -20.98 & -20.82 & -18.94 \\
FCNN  & -21.95 & -18.75 & -18.31 & -16.52 \\
\hline \hline
\end{tabular}
\end{threeparttable}
\end{table}

As we transition into multi-antenna scenarios for UEs, it is imperative to integrate an additional LSTM module subsequent to the FC layers in the decoder. This adaptation is geared towards efficiently extracting correlations existing between proximate antennas at the UE. As a use case, let us consider a setup where both the RIS and the UE are equipped with 64 passive elements and 4 antennas respectively. Table \ref{MAFig} offers a comparison of the feedback performance between standard FC NNs and NNs augmented with LSTM modules when BPD is set at 0.3. For clarity, ``LSTM'' signifies NNs integrated with LSTM modules, while ``FCNN'' denotes the conventional FC NNs. The table encapsulates the enhancement in feedback accuracy across various path numbers, attributed to the integration of LSTM modules.

\begin{figure}[t]
    \centering
     \includegraphics[width=0.9\linewidth]{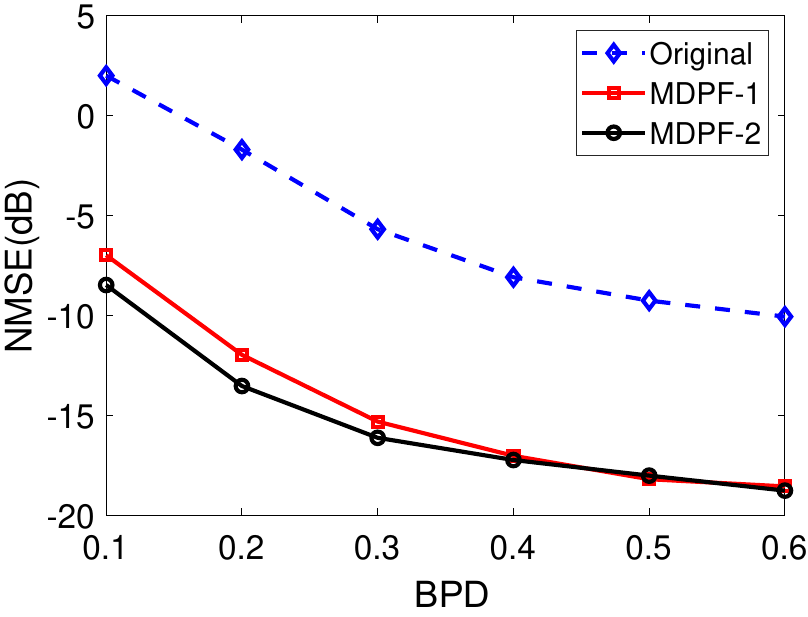}
	\caption{\label{PhaseFig}
	NMSE (dB) performance of the CSI phase feedback. The RIS is equipped with 256 passive elements, the UE is equipped with a single antenna, the first SCM dataset is adopted, and the channel path number is $L=3$.}
\end{figure}

\begin{figure*}[t]
    \centering
    \subfigure []{
     \includegraphics[width=0.45\linewidth]{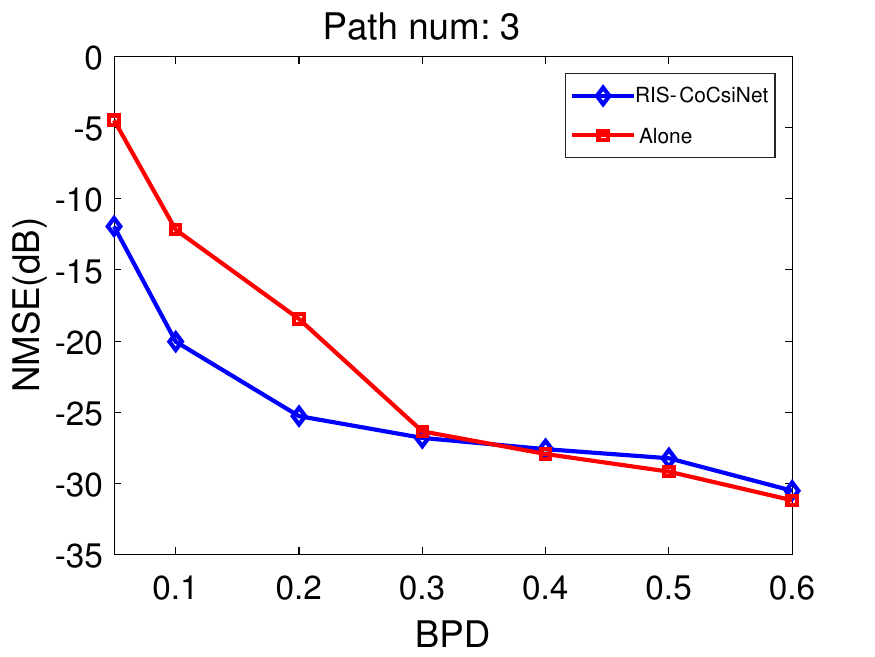}}
	\subfigure[]{
	\includegraphics[width=0.45\linewidth]{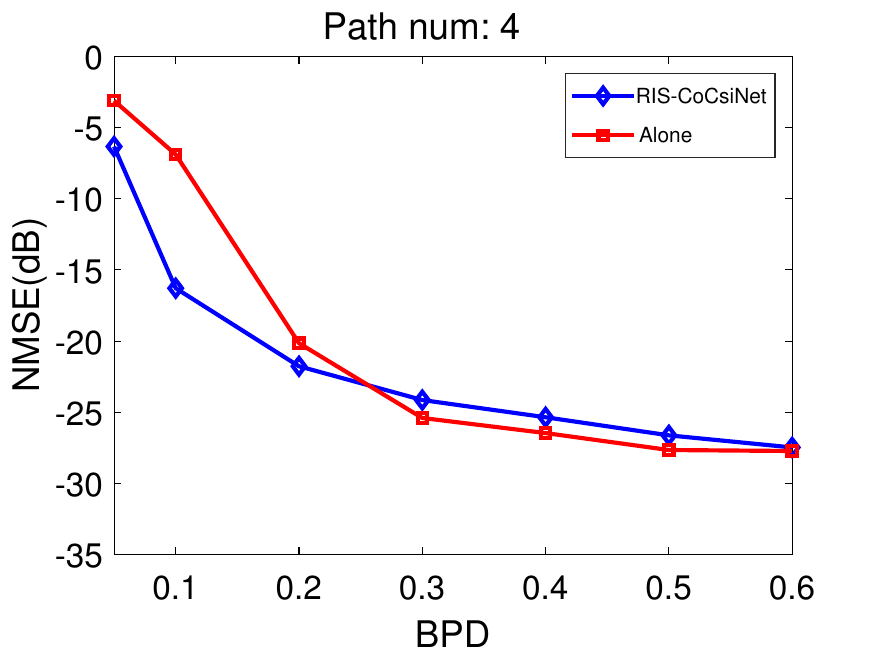}	}
     \subfigure[]{
	\includegraphics[width=0.45\linewidth]{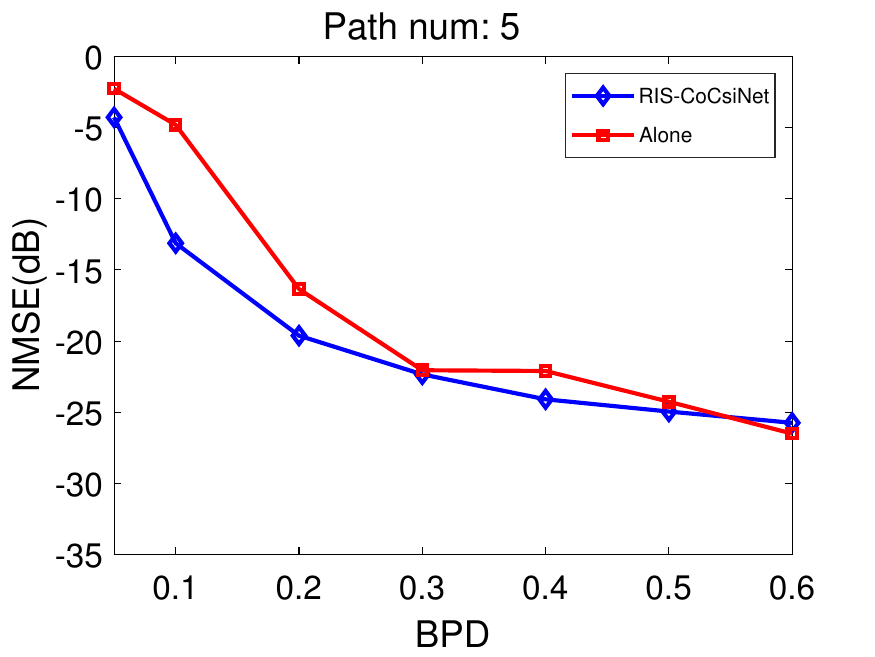}	}
\subfigure[]{
	\includegraphics[width=0.45\linewidth]{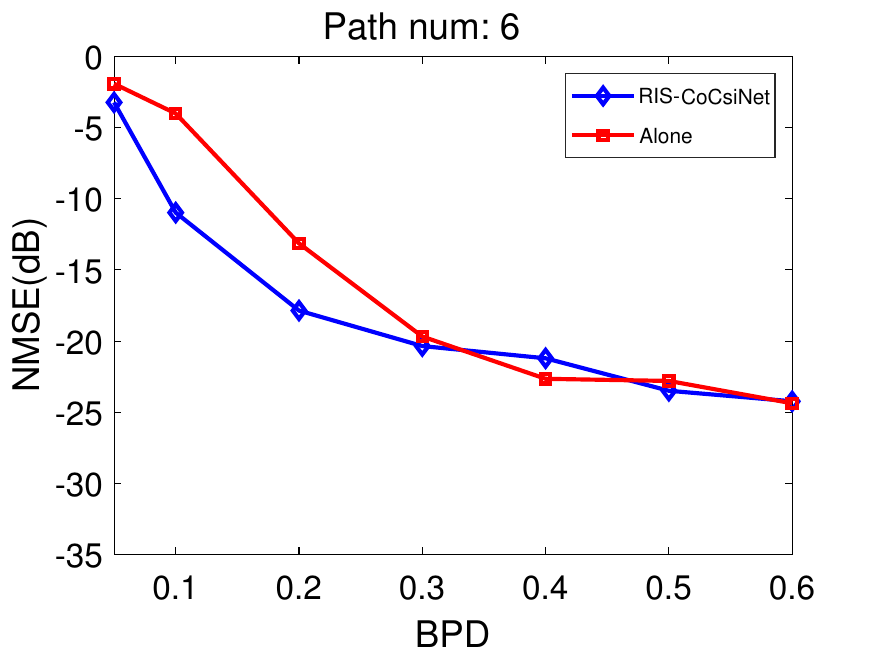}	}
	\caption{\label{coFig}
	NMSE (dB) comparison of RIS-CoCsiNet versus non-cooperative NNs with $N_{\rm s}=256$ and $N_{\rm r}=1$. The first SCM dataset is utilized, and channel path numbers are $L=3,4,5,6$.}
 \vspace{-0.6cm}
\end{figure*}

\subsection{Performance of MDPF NNs}
The focus of this subsection shifts to the evaluation of the MDPF NNs, proposed for enhancing the feedback process, using the first SCM dataset. One significant modification introduced in the MDPF strategy is the incorporation of magnitude information into the loss function, specifically targeting phase feedback.

An intriguing aspect of compressing and feedback for RIS-UE CSI phase lies in its dependence on the CSI magnitude. This relationship suggests that a simple NMSE computation between the original and the reconstructed RIS-UE CSI phase isn't straightforward. Rather, an NMSE determination between the original and the reconstructed complex RIS-UE CSI is required, which is mathematically represented as:
\begin{equation}
\label{NMSE4phase}
{\rm NMSE} = {\rm E} \bigg \{  \frac{ \|  ( \measuredangle {\bf H} -   \measuredangle  \widehat{\bf H}  ) \odot|{\bf H}|  \|_2^2}{ { \Vert   {\bf H} \Vert }_2^2 }   \bigg \}.
\end{equation}

Considering a typical scenario with a 256-element RIS and a single-antenna UE, and a channel path number of $L = 3$, we get a more detailed insight into the effectiveness of the MDPF approach. As illustrated in Fig. \ref{PhaseFig}, the BPD-NMSE trade-offs for MDPF-1, MDPF-2, and the standard phase feedback method are plotted. For extremely low BPDs, the NMSE of the original phase feedback approach surpasses the 0 dB threshold. This indicates that minimal pertinent information about the CSI phase is relayed back. Such an occurrence can be attributed to the NNs' inability to discern vital information. Consequently, these networks might strive to relay all phase details, consequently demanding higher feedback bits. This setback is remarkably alleviated by the MDPF-1 and MDPF-2 strategies, with an impressive improvement of 8.93 dB achieved at a BPD of 0.5.

Furthermore, MDPF-2, with its advantage of leveraging instant CSI magnitude information, exhibits superior performance relative to MDPF-1, especially at lower BPD values. However, as the BPD escalates, this performance delta narrows down, likely because the available feedback bits sufficiently cater to the requisite phase information in both MDPF-1 and MDPF-2.
As the BPD increases, the gap becomes smaller, because the feedback bit is enough for all required phase information for MDPF-1 and MDPF-2.

\subsection{Performance of Cooperative CSI Feedback for RIS-assisted Multi-user Systems}
In this subsection, we evaluate the performance of the proposed cooperative CSI feedback NNs, RIS-CoCsiNet, using both datasets and elucidate the mechanism through NN parameter visualization.

\subsubsection{Comparison between RIS-CoCsiNet, CS, and NNs without Cooperation}
Fig. \ref{coFig} compares the performance of RIS-CoCsiNet to non-cooperative NNs using the first SCM dataset. The setup includes a RIS with 256 elements, a UE with a single-receiver antenna, and channel path numbers $L=3,4,5,6$. With restricted BPD, RIS-CoCsiNet noticeably outperforms its non-cooperative counterpart. This superiority stems from the challenges in CSI recovery at the BS with limited feedback from a single UE. Thus, cooperative feedback becomes advantageous. However, as BPD exceeds 0.3, the performance difference between the two approaches narrows since adequate feedback from the lone UE suffices for precise CSI reconstruction.

\begin{figure}[t]
    \centering
     \includegraphics[width=0.9\linewidth]{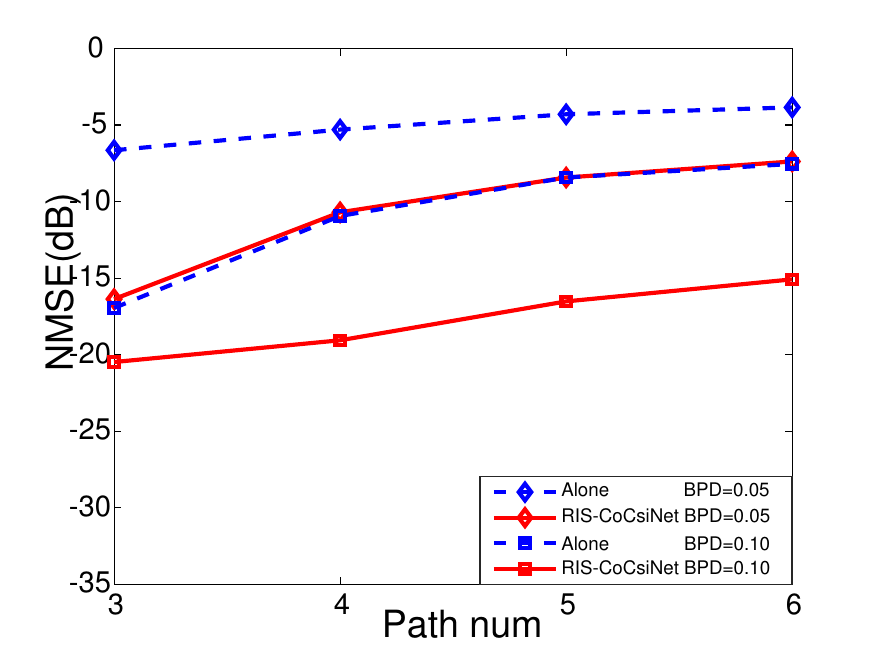}
	\caption{\label{d2dma}
	Comparison of NMSE (dB) between the RIS-CoCsiNet equipped with LSTM modules and the non-cooperative NNs using the SCM dataset. Here, $N=64$, $N_{\rm r}=4$, BPD values are 0.05 or 0.10, and channel path numbers are $L=3,4,5,6$.
}
\vspace{-0.6cm}
\end{figure}

In Fig. \ref{d2dma}, the feedback efficiency of RIS-CoCsiNet with LSTM modules is contrasted against non-cooperative NNs, using the first SCM dataset. This scenario has a RIS with 64 elements and a UE with 4 antennas. The evaluated BPDs are 0.05 and 0.10, with channel path numbers set at $L=3,4,5,6$. At low BPDs, RIS-CoCsiNet significantly excels. Notably, the performance differential varies with increasing channel paths based on BPD values. For a very low BPD of 0.05, RIS-CoCsiNet's advantage reduces with more channel paths, but for a BPD of 0.10, the advantage amplifies.

\begin{figure}[t]
    \centering
     \includegraphics[width=0.9\linewidth]{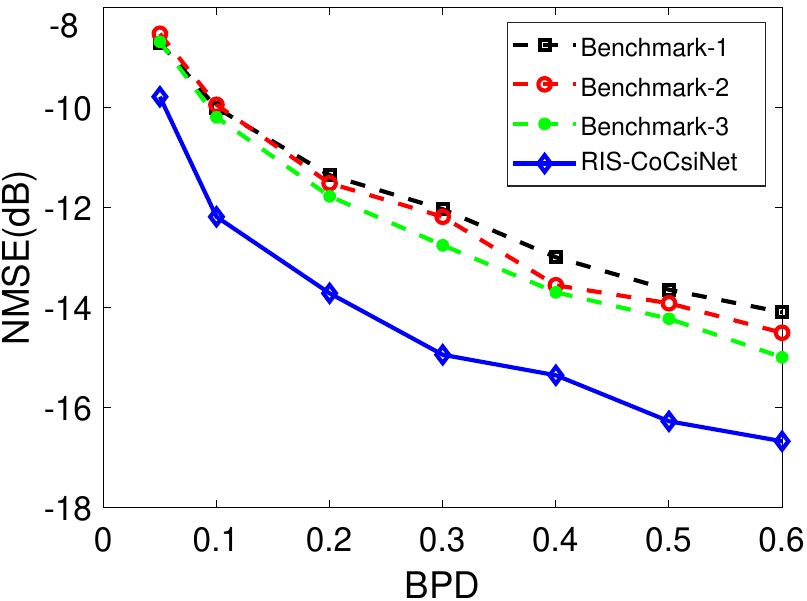}
	\caption{\label{d2dcomplexity}Comparison of NMSE (dB) between the RIS-CoCsiNet and non-cooperative NNs with 256 antennas at the BS, one antenna at the UE, and CSI produced by QuaDRiGa. }
  \vspace{-0.6cm}
\end{figure}

We further validated RIS-CoCsiNet's efficiency by evaluating its cooperative framework on a second channel dataset conforming to 3GPP TR 38.901 v15.0.0 \cite{3gpp2018study}. Unlike the initial comparison focusing only on the baseline NN, this assessment considers multiple benchmarks. This ensures that RIS-CoCsiNet's superior performance is not solely due to NN complexity, as shown in Fig. \ref{d2dcomplexity}.

\begin{figure}[t]
    \centering
     \includegraphics[width=0.9\linewidth]{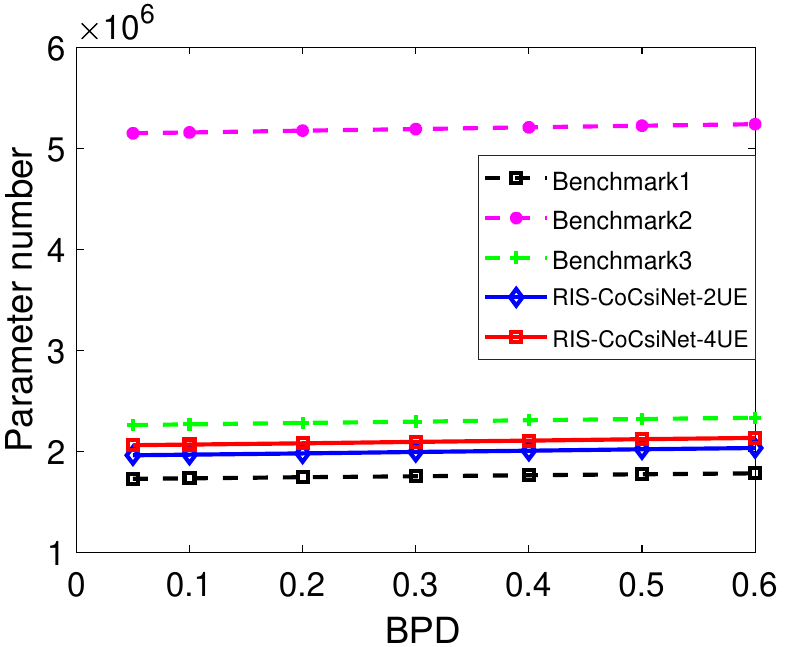}
	\caption{\label{NNnumber} FLOPs count comparison among RIS-CoCsiNet, Benchmark-1, Benchmark-2, and Benchmark-3 NNs.}
 \vspace{-0.6cm}
\end{figure}

Benchmark-1 denotes the baseline architecture previously illustrated in Fig. \ref{coFig}. Benchmark-2 characterizes a network architecture where the neuron count of the FC layers at the decoder is double that of the baseline architecture. Benchmark-3 is designed akin to the proposed RIS-CoCsiNet, but with a twist: while the original co-decoder input was feedback from the cooperative UE, here it takes feedback from a single UE. As the figure suggests, the proposed CoCsiNet surpasses all benchmarks, underlining the significance of leveraging the correlation in the CSI magnitude among proximate UEs. Unlike the results observed with the first dataset, CoCsiNet consistently outshines the non-cooperative NNs regardless of BPD. The disparity in performance can potentially be attributed to variations in channel complexity, as diverse datasets yield different outcomes. As channel complexity escalates, so does the performance gap. In simpler channels, this gap becomes inconsequential at high BPD values.

DL-based feedback methodologies are further juxtaposed with the CS-based technique. Table \ref{CS} contrasts their NMSE performance. The CS problem is addressed using the CVX toolbox. Notably, the CS-based method falls considerably short when pitted against DL-based strategies. This observation mirrors findings from \cite{8322184}, especially when dealing with a low compression ratio. Within the scope of this work, the compression ratio for the CS-based approach is markedly low. For instance, with a BPD of 0.1, the ratio sits at 1/40. The RIS-UE CSI undergoes an initial compression by a factor of 40, followed by a discretization via a 4-bit uniform quantizer. The inherent limitations of CS become apparent in such scenarios, given that it solely relies on sparsity as its prior information, neglecting the nuances of codeword quantization.

Furthermore, Table \ref{CS} pits RIS-CoCsiNet against varying numbers of cooperative UEs. A scenario with four cooperating users exhibits marginally superior performance to the two-UE setup. This is attributed to the fact that, as the number of cooperative UEs increases, shared information demands fewer feedback bits per UE. A notable deviation emerges when BPD is 0.05, where the two-user cooperative setup trumps the four-user counterpart. This stems from the increasing challenge in training NNs with a rising UE count, a difficulty exacerbated at low BPDs.

\begin{table*}[t]
\centering
\caption{NMSE ($\rm dB$) performance comparison (3GPP\_38.901 channel).}
\label{CS}
\begin{tabular}{c|cccccccccccc}
\hline \hline
 \diagbox{Method}{BPD}    & 0.05   & 0.1    & 0.2    & 0.3    & 0.4    & 0.5    & 0.6    & 0.7   & 0.8   & 0.9   & 1.0   & 2.0   \\  \hline\hline
CS           & 0.65   & 0.76   & 0.74   & 0.60   & 0.38   & 0.13   & -0.12  & -0.46 & -0.85 & -1.26 & -1.66 & -6.43 \\
Alone        & -8.71  & -10.01 & -11.36 & -12.04 & -13.00 & -13.65 & -14.10 & /     & /     & /     & /     & /     \\
RIS-CoCsiNet-2UE & -10.62 & -11.31 & -13.57 & -14.98 & -15.43 & -15.82 & -16.38 & /     & /     & /     & /     & /     \\
RIS-CoCsiNet-4UE & -9.79  & -12.19 & -13.72 & -14.95 & -15.36 & -16.28 & -16.68 & /     & /     & /     & /     & /  \\     \hline\hline
\end{tabular}
\end{table*}

\subsubsection{NN Complexity Analysis}

The NNs discussed in this study predominantly comprise FC layers. As indicated in \cite{molchanov2016pruning}, the floating point operations (FLOPs) count for an FC layer is roughly twice the number of its parameters. Consequently, our focus is directed towards the computation of this parameter count. Fig. \ref{NNnumber} offers a comparative analysis of these counts. The growth in parameter count, when juxtaposed with Benchmark 1, is minimal. Further, the parameter count of RIS-CoCsiNet is found to be less than that of both Benchmarks 2 and 3. Importantly, in the proposed RIS-CoCsiNet, additions at the BS are confined to the co-decoder and the combination NN, with the UE side remaining unchanged. As highlighted in \cite{guo2019convolutional}, while UEs operate under constraints (limited memory, computational capabilities, and battery life), the BS is resource-rich. As a result, the marginal increase in complexity attributed to the proposed CoCsiNet is inconsequential. Furthermore, when benchmarked against existing studies, such as \cite{9931713}, the complexity of our NN is considerably reduced. Specifically, the FLOP count for the proposed RIS-CoCsiNet stands at around 4\,M, in contrast to the majority of prior work, which exceeds 10\,M FLOPs (a more detailed comparison is available in \cite[Fig. 30]{9931713}). This underscores the modest computational demands of our approach.

\subsubsection{Cooperation Mechanism}
\begin{figure*}[t]
    \centering
      \subfigure [Encoder parameter visualization for UE 1]{
     \includegraphics[width=0.45\linewidth]{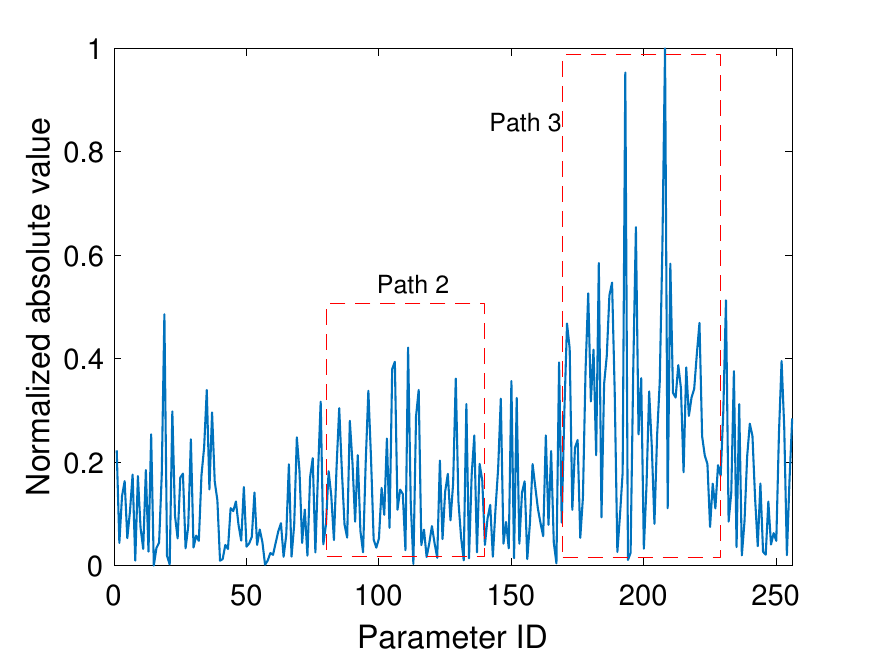}}
	\subfigure[Encoder parameter visualization for UE 2]{
	\includegraphics[width=0.45\linewidth]{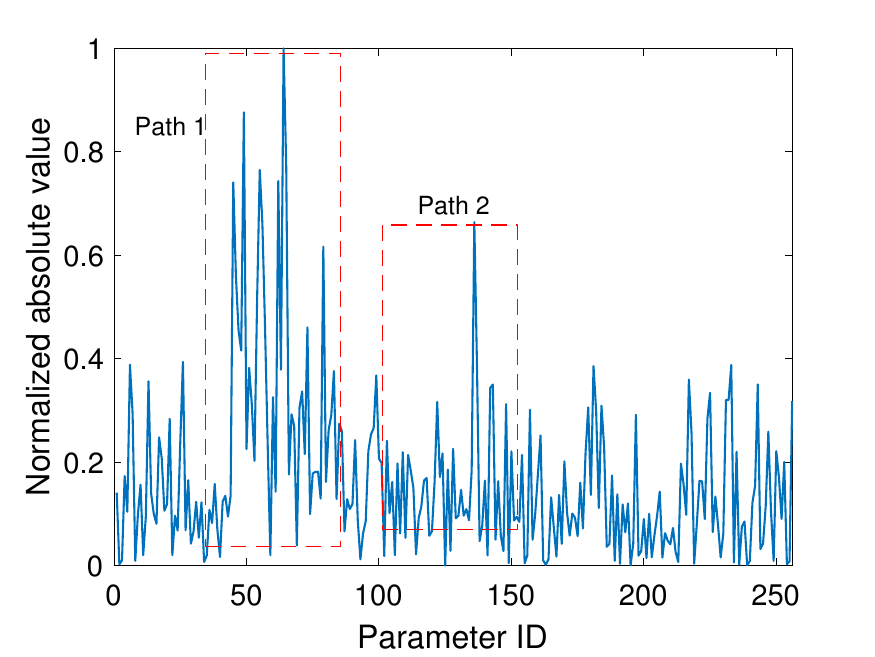}	}
	\caption{\label{FCFigD2D} Visualization of the row-sum norm from the first FC layer of encoders in the RIS-CoCsiNet using the first SCM dataset, with a BPD of 0.05 and channel path number $L=3$. Higher absolute values indicate areas of increased focus for the NNs. }
  \vspace{-0.6cm}
\end{figure*}

A prime incentive for pinpointing NN attention through parameter visualization emerges from the realm of NN pruning. Here, connections with smaller weights are typically pruned \cite{guo2019compression}. Empirical findings from \cite{guo2019compression} indicate that feedback is intimately tied to high-weight connections, which can be conceptualized as focal areas. Utilizing this parameter visualization approach, researchers in \cite{guo2019convolutional} discerned that the CsiNet+ proposed in \cite{guo2019convolutional} concentrates primarily on non-sparse zones.

Outlined in Table \ref{NNArchitecture}, the dimension of the first FC layer at the encoder is defined as $N_{\rm r}N_{\rm s}\times 2N_{\rm r}N_{\rm s}$. The row-sum norm of the weights is obtained using:
\begin{equation}
W_{i} = \sum_{j=1}^{2N_{\rm r}N_{\rm s}} |W_{i,j}|,
\end{equation}
where $W_{i,j}$ signifies the weight linking the $i$-th input and $j$-th output neurons.
Fig. \ref{FCFigD2D} illustrates the row-sum norm vector, $[W_{1},\ldots, W_{N_{\rm r}N_{\rm s}}]$, for two encoders in the RIS-CoCsiNet. The RIS has 256 passive components, the UE uses a single antenna, the primary SCM dataset is applied, and both the BPD and the channel path number are set at 0.05 and 3, respectively. From the RIS-UE CSI data, three main paths are evident around Parameter IDs 50, 125, and 200, highlighted with dotted rectangles. The first encoder focuses on the third and the second paths, while the second targets the first and the second paths. Together, these encoders transmit information from different paths.
Without the aid of the cooperation strategy, each encoder needs to focus on all paths and transmit the information of all paths.
Fortunately, shared information in RIS-CoCsiNet no longer needs to be transmitted repeatedly, and the saved overhead can be used to help more accurate feedback on individual information.
In addition, during training, the NNs have learned the CSI distribution and potential correlation among nearby UEs.
Thus, the encoders in the proposed cooperation strategy can automatically focus on different parts of the RIS-UE CSI without the aid of D2D communications.
This efficient mechanism is why cooperative NNs often outperform non-cooperative ones.

\section{Conclusion}
\label{s6}

This study introduced a deep learning-based cooperative CSI feedback and recovery system, dubbed RIS-CoCsiNet, designed to harness the RIS-UE CSI correlation among proximate UEs in RIS-supported multi-user environments. The CSI information from these adjacent UEs is bifurcated into shared and unique segments. In a novel approach, an additional decoder alongside a combination NN is integrated at the BS to process the shared information obtained from feedback of two nearby UEs, while individual data is handled by each UE's personal decoder. Subsequently, this amalgamation of data takes place at the BS through FC layers.
A distinctive feature of the proposed RIS-CoCsiNet is its ability to function without the necessity for CSI sharing, thereby eliminating the need for inter-device communication. This work also delved into two predominant bit generation techniques --- quantization and binarization --- and showcased two MDPF frameworks that infuse CSI magnitude intel into the CSI phase feedback.
Empirical results evinced the superior efficacy of RIS-CoCsiNet over non-cooperative NNs, particularly when feedback overheads are tightly restricted. An NN parameter visualization provided insights into the cooperative methodology. Within the RIS-CoCsiNet system, encoders from two close-by UEs collaboratively siphon information from varied channel paths, which substantially curtails redundant feedback overhead associated with shared information.

\ifCLASSOPTIONcaptionsoff
  \newpage
\fi

\bibliographystyle{IEEEtran}
\bibliography{IEEEabrv,reference}

\end{document}